\newcommand{\gtrsim}{\lower 2pt \hbox{$\, \buildrel {\scriptstyle >}\over {\scriptstyle\sim}\,$}}

\documentclass[
    ,final            
  ]
  {aipproc}

\layoutstyle{6x9}

\begin{document}

\title{Wide Area X-ray Surveys for AGN and Starburst Galaxies}

\classification{98.54.Ep, 98.62.Ve, 98.70.Qy}
\keywords      {galaxies, active galactic nuclei, x-rays}

\author{Andrew Ptak}{
  address={Johns Hopkins University, Department of Physics and Astronomy}
}

\begin{abstract}
While often the point sources in X-ray surveys are dominated by AGN,
with the high sensitivity of modern X-ray telescopes such as Chandra
and XMM-Newton normal/starburst galaxies are also being detected in
large numbers.  We have made use of Bayesian statistics for both the
selection of galaxies from deep X-ray surveys and in the analysis of
the luminosity functions for galaxies.  These techniques can be used
to similarly select galaxies from wide-area X-ray surveys and to
analyze their luminosity function.  The prospects for detecting
galaxies and AGN from a proposed ``wide-deep'' XMM-Newton survey and
from future wide-area X-ray survey missions (such as WFXT and eRosita)
are also discussed.
\end{abstract}

\maketitle

\section{X-ray Surveys}
Extragalactic X-ray surveys are a powerful tool to study important source
populations such as active galactic nuclei (AGN), 
clusters of galaxies, and, recently, normal galaxies
\cite{BrandtReview}.  In the 
bandpasses of Chandra and XMM-Newton, X-rays penetrate column
densities up to $10^{23-24}\rm\ cm^{-2}$, and are therefore efficient at
detecting moderately-obscured ``Compton-thin'' AGN.  Compton-thick AGN
(AGN with column densities of $>10^{23-24}\rm\ cm^{-2}$) can also be
detected in X-rays when prominent scattered emission is present
(typically of order of $\sim 1\%$ of the intrinsic emission) \cite{Cappi06}
and can be detected in 
hard (E $>$ 10 keV) surveys \cite{Winter2008}.  Therefore X-ray surveys are
essential for 
 a complete census of AGN.  Normal galaxies can now be detected in
 large numbers thanks to the high sensitivity of Chandra and
 XMM-Newton, however since they have low luminosities ($L_X < 10^{42}
 \ \rm ergs\ s^{-1}$), they are more difficult to detect than AGN.

\subsection{X-rays from Normal/Starburst Galaxies}
It has been known since the early 1980s that the X-ray emission of
normal and starburst galaxies (galaxies with very high star-formation
rates) are correlated with both the star-formation rate and stellar
mass of the galaxies \cite{Fabbiano89}.
The physics behind this is that high-mass stars evolve rapidly (on
time scales of $10^{6-8}$ years), and in turn explode as supernovae
(SN).  Occasionally these SN are detected in X-rays, however more
often they heat the ISM to X-ray emitting temperatures (i.e., T $>
10^{6-7}$ K) and produce neutron stars and black holes in X-ray
binaries.  X-ray binaries where the companion is a high-mass star
(high-mass X-ray binaries) also have short evolutionary time scales.
Therefore hot ISM and high-mass X-ray binaries track the current
star-formation rate.  Low-mass X-ray binaries have longer evolutionary
time scales (on the order of Hubble times), and therefore track the
integrated star-formation history of galaxies (i.e., the total
stellar mass).

\subsection{X-ray Galaxy Survey Strategies}
There are several approaches to surveying specific sources types with
low fluxes such as X-ray observations of galaxies.
Deep, pencil-beam surveys of course probe to the faintest fluxes
however due to the limited survey volume tend to result in low numbers
of rare, high-luminosity objects.  Wide area surveys (e.g., XMM-COSMOS)
detect large numbers of AGN but are usually too shallow or survey too
small of an area to detect significant numbers of galaxies. Another approach
is to correlate large catalogs (e.g., the RC3 catalog or the SDSS)
with archival data.  For example, $\gtrsim 400$ galaxies have been
detected in X-rays based on correlating the SDSS with the 2XMM catalog
(see Parnau et al. these proceedings).  We are pursuing this approach
in the case of {\it Chandra} and {\it XMM-Newton} archival data by
taking advantage of the 
XAssist pipeline processing of these data\footnote{see
\url{http://www.xassist.org}} and correlating the fields with the RC3
catalog.  Note that by working with the original data rather than
simply a source catalog we will be properly integrating the X-ray flux
over the full RC3 ellipse for each galaxy, which is often larger that
the telescope PSF, and also will be able to compute upper-limits for
galaxies not detected in X-rays. 

\begin{figure}
  \includegraphics[height=.4\textheight]{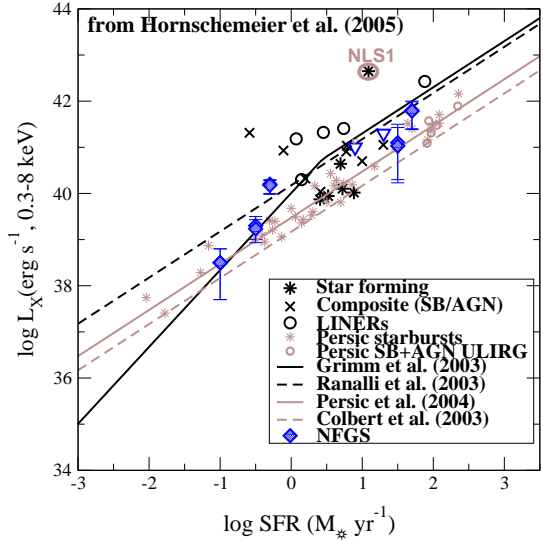}
  \caption{The X-ray/SFR correlation based on correlating SDSS
    galaxies and the Chandra archive \cite{HornSDSS}, with
    data from XMM observations of the Nearby Field Galaxy Survey
    (NFGS) added. \label{nfgs-plot}}
\end{figure}

Finally we discuss the option of pursuing X-ray observations of
statistically-complete samples that are not X-ray selected.  We have
been observing galaxies selected from the Nearby Field Galaxy Survey,
which have very-well determined star formation 
rates \cite{Kewley02}.  We have received six XMM-Newton datasets
and also included one serendipitous Chandra observation
\cite{Ptak2008}.  The X-ray/SFR correlation based on correlating the
Chandra archive with the Kauffman SDSS
galaxy catalog \cite{Kauff2003mass} is shown in Figure
\ref{nfgs-plot}\cite{HornSDSS},  
with the X-ray NFGS data also plotted.  The NFGS points are consistent with
either the lower X-ray/SFR normalization implied by the X-ray detected
SDSS galaxies, or a break in the X-ray/SFR correlation.  Clearly a
larger unbiased sample is needed.

 
\section{Bayesian Selection and Analysis of Normal Galaxies in Deep
  X-ray Surveys}
In \cite{Norman2004} and \cite{Ptak2007}, normal/starburst
galaxies were selected from the Chandra Deep Fields (CDF) using a Bayesian
model selection methodology.  Our motivation for employing this
technique was to {\it directly take into account the X-ray measurement
  errors} which can be significant for X-ray sources at the flux limit
of the survey.

\subsection{Galaxy Selection}
The most discriminating features for separating galaxies from AGN were
found to be X-ray hardness ratio, (H-S)/(H+S), where H and S are the
numbers of photons above and below 2 keV, respectively, X-ray
luminosity, and X-ray/optical flux ratio.  In our analysis of GOODS data we also
included the X-ray/near-IR flux ratio.  We determined 
``parent'' distributions for these parameters by selecting a sample of
normal galaxies, type-1 AGN and type-2 AGN from the CDF South based on
high quality optical spectroscopy, and then taking the mean and
standard deviation ($\sigma$) for each parameter.  

The posterior probability for observing the parameters $\theta$, where
here 
$\theta = \{HR, \log L_X, \log F_X/F_{\rm opt}, \log F_X/F_{\rm
NIR}\}$ given the data $D$ is given by Bayes' theorem: 
$p_M(\theta|D) = p_M(\theta)p_M(D|\theta)/p_M(D)$.
The $M$ subscripts denote that this is assuming a given model $M$,
here galaxies, type-1 AGN (AGN1) and type-2 AGN (AGN2).  If
multiple models are being considered, then the prior probability for
each model must also be included.  $p_M{\theta}$ are the ``prior'' distributions for the parameters for a 
given model $M$, for which we used the parent distributions discussed
above. $p_M(D|\theta)$ is the likelihood function of observing the
parameters $\theta$ given the data.  Often $p_M(D)$ is
considered to be a normalization constant since it does not depend on
the model parameters $\theta$, and is defined to be 
$p_M(D) = \int d\theta p_M(\theta|D)$.
However $p_M(D)$ is also often (perhaps more precisely) considered to be
the marginal likelihood or Bayesian evidence for the model $M$.  The
relative probability of two competing models given the data is then
given by the ``Bayes Factor'' or
\begin{equation} 
\frac{p_{M_1}(D)}{p_{M_2}(D)} = \frac{\int d\theta p_{M_1}(\theta|D)}{\int d\theta p_{M_2}(\theta|D)}
\end{equation}
Sources with $p_{\rm galaxy}/p_{\rm AGN1} > 1$ and $p_{\rm
  galaxy}/p_{\rm AGN2} > 1$ were selected as galaxies\footnote{In
  \cite{Ptak2007}, a more conservative sample was selected with Bayes
  factors $>3$.}.
Here we are assuming a flat prior on the numbers of sources for each
model $M$, in other words we are assuming that
the ``true'' number of normal/starburst galaxies, type-1 AGN and type-2 AGN are
approximately equal.  

\subsection{Evolution}
\begin{figure}
\includegraphics[height=0.3\textheight]{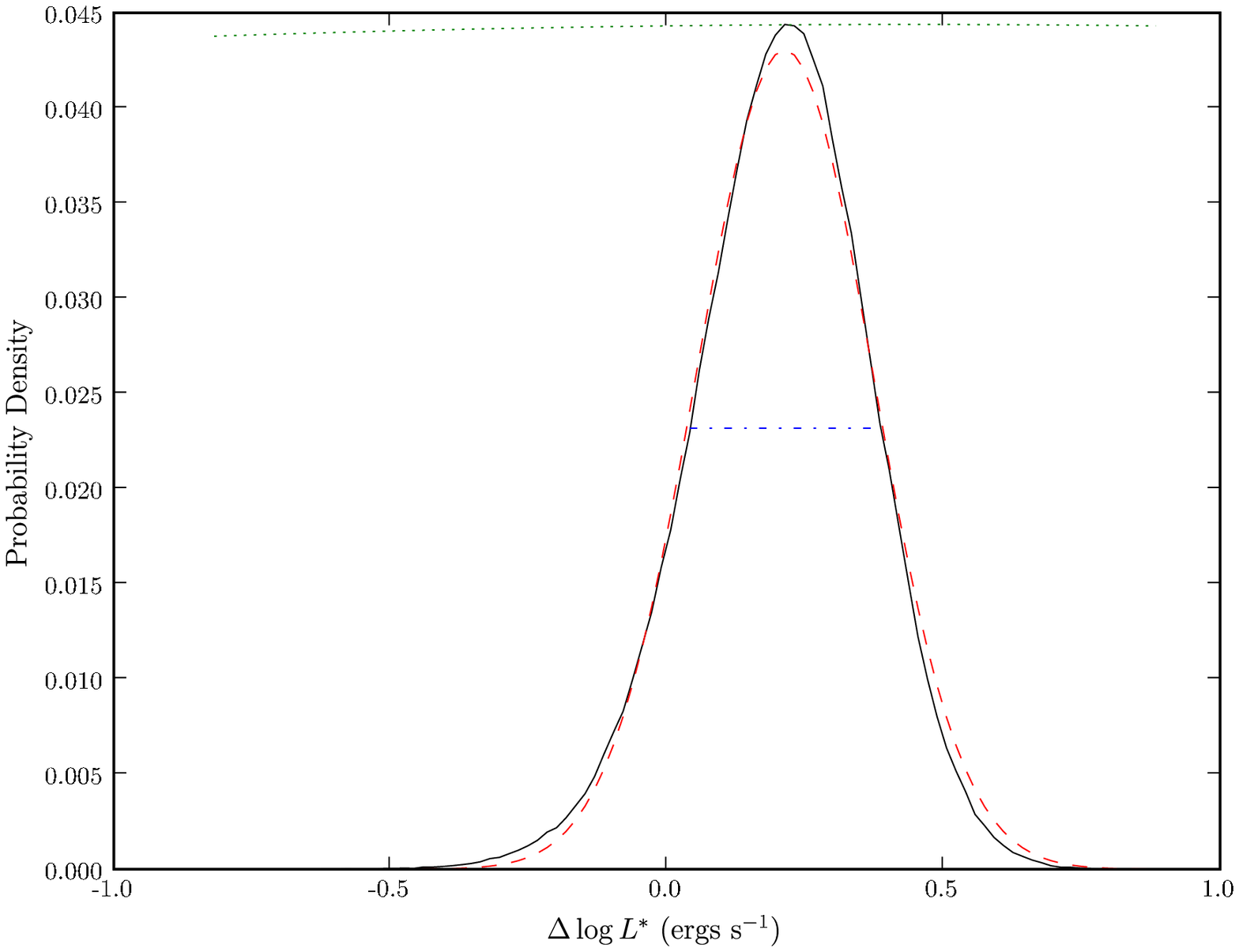} 
\includegraphics[height=0.3\textheight]{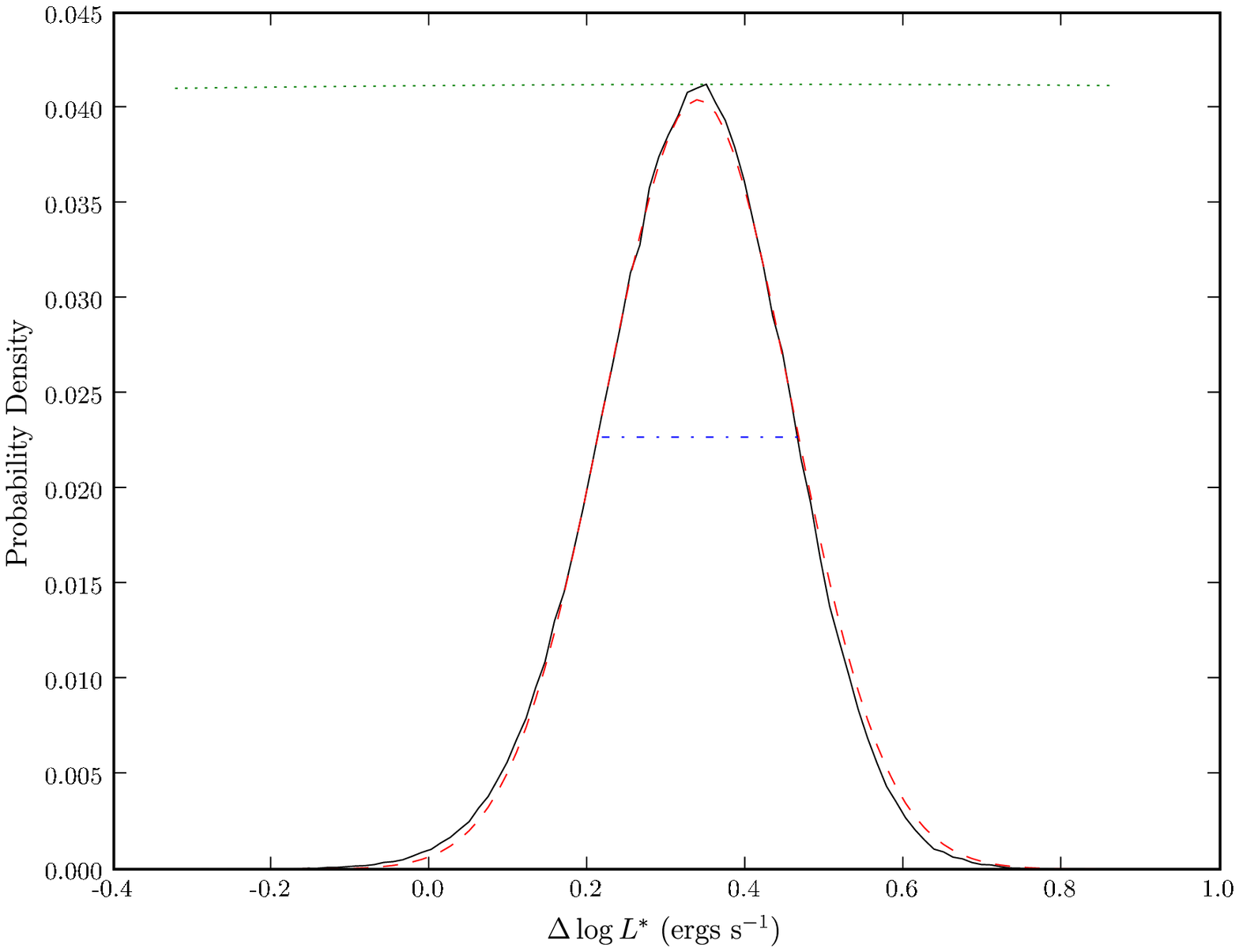}
\caption{Posterior probability for the change in $\log L^*$ between the
  $z \sim 0.25$ and $z\sim 0.75$ for early-type (left) and late-type
  (right) galaxies.  The solid (black) lines show the posterior probabilities,
  while the dashed (red) lines show Gaussian distributions with the
  same mean and standard deviation as the posteriors, and the dotted
  (green) lines show the prior distributions ($\sim$ flat in these
  cases)
\label{post}} 
\end{figure}

Evolution was observed qualitatively 
between the redshifts of 0.25 and 0.75 consistent with pure luminosity
evolution, $L^*(z) = L^*(z=0)(1+z)^p$ with $p \sim 3$ by comparing the
X-ray luminosity functions (XLFs) with the far-infrared luminosity
functions \cite{Norman2004}.  Subsequently using GOODS data we fit the X-ray luminosity
functions by using Markov-Chain Monte Carlo (MCMC) \cite{Ptak2007}. 
MCMC analysis results in a
distribution of parameter values (the ``chains'').  This allows
the direct visualization of posterior probabilities for important
quantities, such as the change in $L^*$ between the low and high redshift
XLFs (Figure \ref{post}). Another key advantage of the Bayesian approach is that ``derived'' quantities such as luminosity density can be computed 
directly from the chain parameter values, allowing the posterior
probabilities for these quantities to be visualized or summarized
(i.e., with the mode as a ``best-fit'' value and the 68\% confidence
interval as the ``error bar'') without questionable propagation of
error \cite{Ptak2007,Kelly2008}.  In future work we will be
incorporating the additional 1 Ms of CDF-S data, as well as improving
our MCMC analysis of the XLFs to also include VLA radio and Spitzer
mid-IR data (both of which are star-formation rate indicators and will
help discriminate galaxies from AGN). 

\section{Wide-Area X-ray Surveys}
\subsection{XMM Wide-Area Surveys}
\begin{figure}
\includegraphics[height=0.35\textheight]{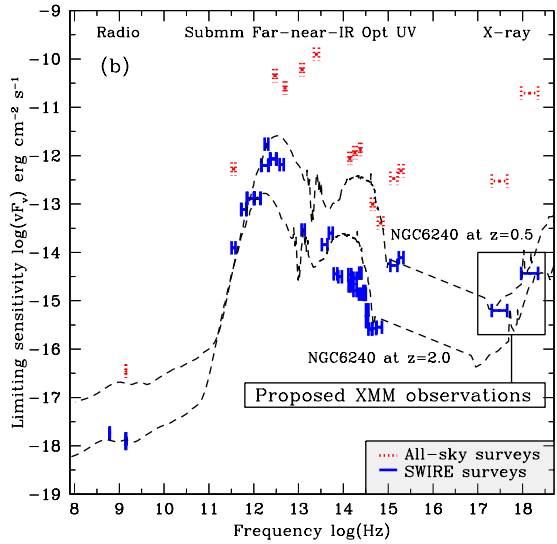}
\caption{
The SED of NGC 6240 (dashed curves, for redshifts 
  of 0.5 and 2.0 and assuming a $L_X = 5 \times
  10^{43}\rm\ ergs\ s^{-1}$) shown with the limiting
  fluxes of the SWIRE survey (blue), all-sky sureys (red) and the
  proposed XMM-Newton survey marked.
\label{xmm-wide-deep}}
\end{figure}

While both XMM-Newton and Chandra are very sensitive X-ray telescopes,
XMM-Newton has a larger field-of-view (FOV) than Chandra and is
therefore more adept at wide-area surveys.  While the total solid
angle of 2XMM is large (hundreds of square degrees), its coverage is
non-uniform and potentially biased since the field
selection is not random.  XMM-COSMOS is a wide-area survey covering 2
deg.$^2$ at an exposure of 40 ks per field, or a limiting point-source 
sensitivity of $\sim 5-10 \times 10^{-16} \rm
\ ergs\ s^{-1}\ cm^{-2}$\cite{Salvato2008}.  Here we briefly discuss a
proposal to extend XMM-COSMOS out to 10 deg.$^{2}$ at a similar mean
exposure per field\footnote{Submitted in the XMM AO-8 proposal round,
  PI David Alexander}.  
This
level of sensitivity would be sufficient to detect the (scattered)
X-ray flux
from Compton-thick AGN with a spectral energy distribution similar to
NGC 6240 at z $\sim 0.5-2.0$ (see Figure \ref{xmm-wide-deep}).  This survey 
should result in 6000-8000 AGN being detected, $\sim 1000$ with at
least 100 photons (sufficient for crude X-ray color analysis) and
$\sim 400$ with at least 300 counts (sufficient for spectral analysis,
including the detection of Fe-K lines).  This survey would also detect
$\sim 150-200$ AGN at z$>$3 ($\sim 15-20$ at z$>$4).  We expect up
$\sim 300$ normal/starburst galaxies would be detected, based on the Ranalli et
al. logN-logS\cite{Ranalli2005}.  The field 
selection is in the Spitzer-SWIRE area, and the proposed X-ray data along
with large amount of ancillary data from other wavebands available in
these fields will allow us to study the coeval evolution of AGN and
star formation 
over a wide range of redshift, environment and luminosity.  

 
\subsection{Future Missions}
\begin{figure}
\includegraphics[height=0.3\textheight]{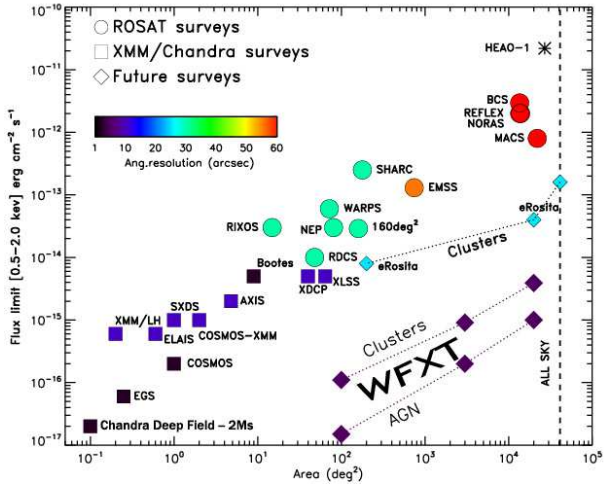}
\caption{The limiting flux and solid angle of the surveys from the
  future missions WFXT and eRosita, where both the sensitivity to AGN
  (point sources) and clusters are shown, along with other X-ray
  surveys.\label{wfxt-flux-area}
}
\end{figure}

\begin{figure}
\includegraphics[height=0.3\textheight]{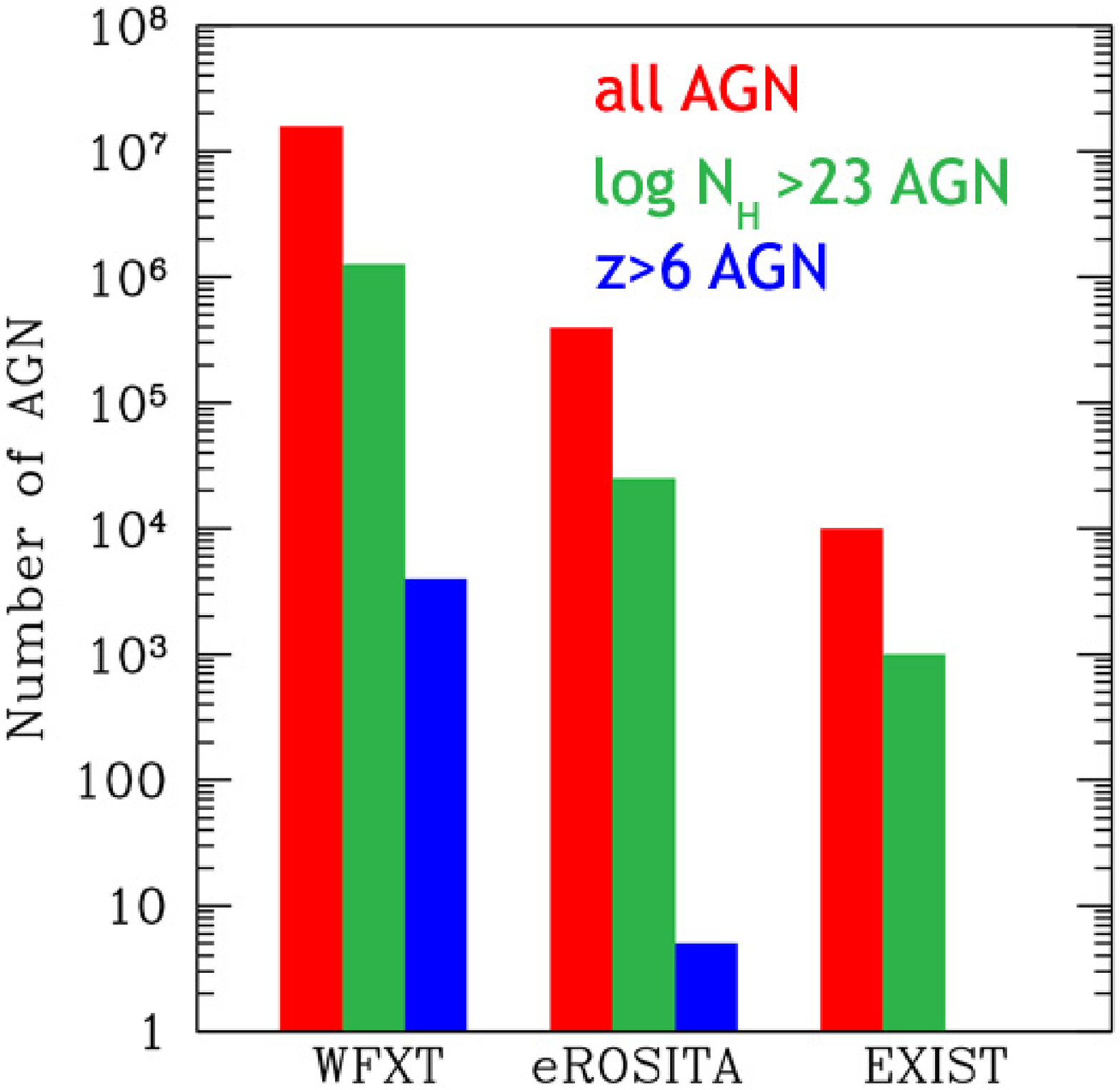}
\includegraphics[height=0.3\textheight]{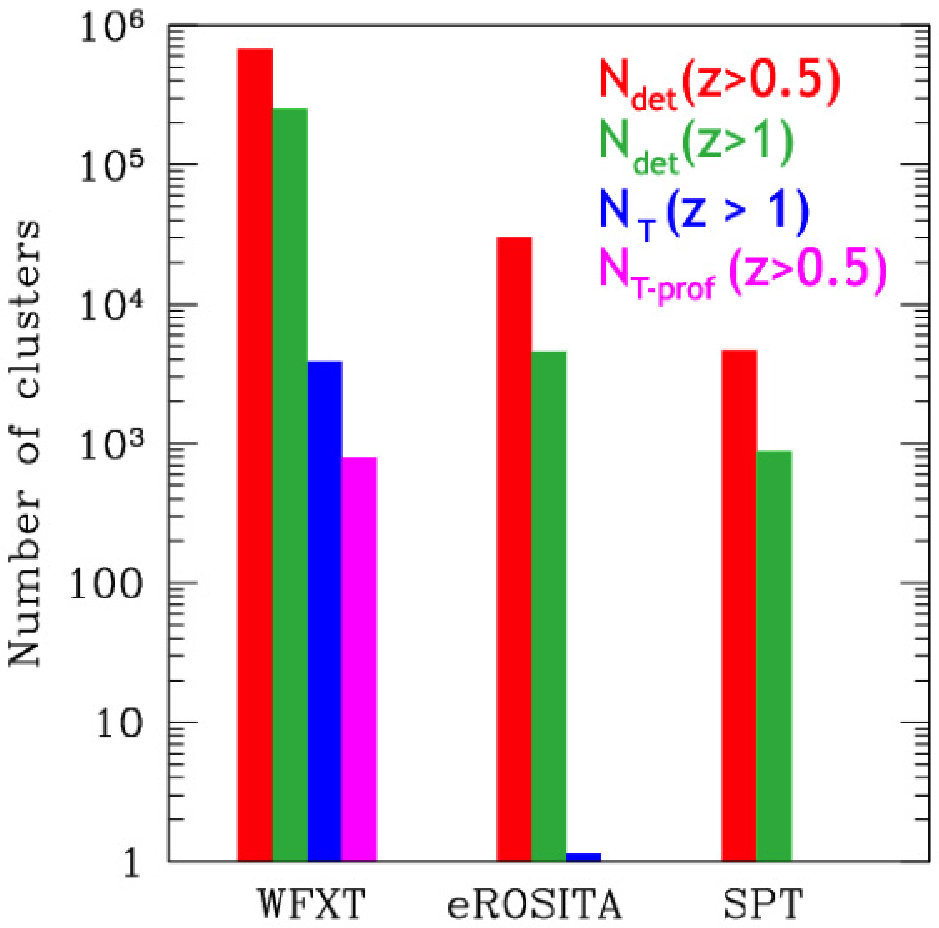}
\caption{Expected numbers of AGN (left) and cluster of galaxies (right) to be
  detected by WFXT and eRosita.  Also shown are the number of clusters
  of galaxies expected to be detected by the South Pole Telescope
  (SPT) via the Sunyaev-Zel'dovich effect.\label{wfxt-numbers}
}
\end{figure}

There are several proposed future missions dedicated to X-ray surveys.
eRosita (extended Roentgen Survey with an Imaging Telescope Array) is
an approved mission expected to fly aboard the Spectrum 
X-Gamma Mission, although the launch date appears to be
uncertain\footnote{http://www.mpe.mpg.de/projects.html\#erosita}.
eRosita will perform several surveys, including a nearly all-sky
shallow survey.  The Wide Field X-ray Telescope (WFXT) is a proposed
medium-class NASA mission to similarly perform several surveys.  The
limiting flux and survey solid angle for both the WFXT and eRosita
surveys are plotted in Figure \ref{wfxt-flux-area}.  The numbers of
AGN and clusters expected to be detect by eRosita and WFXT are shown
in Figure \ref{wfxt-numbers}.  We expect on the order of $\sim 10^4$
normal/starburst galaxies to be detected in the eRosita surveys while
$\sim 10^5$ galaxies should be detected in the WFXT surveys.  Clearly either mission would
drastically increase the numbers of X-ray detected sources and would
be revolutionary.  The high numbers of sources expected from WFXT is
due to both a higher effective area and a smaller PSF (half-energy
width of $\sim 7"$ for WFXT compared to the field-averaged PSF of
$\sim 25-30"$ in the case of eRosita).  WFXT also have the advantage of
using a wide-field optical design \cite{Burrows1992}, giving more uniform
PSF and response across the 1 deg. FOV.


\begin{theacknowledgments}
  We acknowledge the support of NASA grants NNG04GE13G,  NNG05GP14G, and
  NNG06GE59G.
\end{theacknowledgments}



\bibliographystyle{aipproc}   

\bibliography{ms}

\begin{thebibliography}{14}
\expandafter\ifx\csname natexlab\endcsname\relax\def\natexlab#1{#1}\fi
\providecommand{\enquote}[1]{``#1''}
\expandafter\ifx\csname url\endcsname\relax
  \def\url#1{\texttt{#1}}\fi
\expandafter\ifx\csname urlprefix\endcsname\relax\def\urlprefix{URL }\fi
\providecommand{\eprint}[2][]{\url{#2}}

\bibitem[{Brandt} and {Hasinger}(2005)]{BrandtReview}
W.~N. {Brandt}, and G.~{Hasinger}, \emph{\araa} \textbf{43}, 827--859 (2005).

\bibitem[{Cappi} and {et al.}(2006)]{Cappi06}
M.~{Cappi}, and {et al.}, \emph{\aap} \textbf{446} (2006).

\bibitem[{Winter} et~al.(2008)]{Winter2008}
L.~{Winter}, R.~{Mushotzky}, C.~S. {Reynolds}, and J.~{Tueller}, \emph{\apj,
  submitted}  (2008), astro-ph/0808.0461.

\bibitem[{Fabbiano}(1989)]{Fabbiano89}
G.~{Fabbiano}, \emph{\araa} \textbf{27}, 87--138 (1989).

\bibitem[{Kewley} et~al.(2002)]{Kewley02}
L.~J. {Kewley}, M.~J. {Geller}, R.~A. {Jansen}, and M.~A. {Dopita}, \emph{\aj}
  \textbf{124}, 3135--3143 (2002).

\bibitem[{Ptak} et~al.(2008)]{Ptak2008}
A.~{Ptak}, T.~{Heckman}, C.~{Norman}, A.~{Hornschemeier}, L.~{Kewley}, and
  A.~{Zezas}, \enquote{{The X-ray/SFR connection from X-ray observations oft
  the nearby field galaxy sample},} in \emph{"ESAC faculty workshop on x-rays
  from nearby galaxies"}, 2008, pp. 81--84.

\bibitem[{Kauffmann} and {et al.}(2003)]{Kauff2003mass}
G.~{Kauffmann}, and {et al.}, \emph{\mnras} \textbf{341}, 33--53 (2003).

\bibitem[{Hornschemeier} et~al.(2005)]{HornSDSS}
A.~E. {Hornschemeier}, T.~M. {Heckman}, A.~F. {Ptak}, C.~A. {Tremonti}, and
  E.~J.~M. {Colbert}, \emph{\aj} \textbf{129}, 86--103 (2005).

\bibitem[{Norman} and {et al.}(2004)]{Norman2004}
C.~{Norman}, and {et al.}, \emph{\apj} \textbf{607}, 721--738 (2004).

\bibitem[{Ptak} et~al.(2007)]{Ptak2007}
A.~{Ptak}, B.~{Mobasher}, A.~{Hornschemeier}, F.~{Bauer}, and C.~{Norman},
  \emph{\apj} \textbf{667}, 826 (2007).

\bibitem[{Kelly} et~al.(2008)]{Kelly2008}
B.~{Kelly}, X.~{Fan}, and M.~{Vestergaard}, \emph{\apj} \textbf{682}, 874
  (2008).

\bibitem[{Salvato} and {et al.}(2008)]{Salvato2008}
M.~{Salvato}, and {et al.}, \emph{\apj in press}  (2008),
  \urlprefix\url{http://arxiv.org/abs/0809.2098}, astro-ph/0809.2098,
  \eprint{0809.2098}.

\bibitem[{Ranalli} et~al.(2005)]{Ranalli2005}
P.~{Ranalli}, A.~{Comastri}, and G.~{Setti}, \emph{\aap} \textbf{440}, 23--37
  (2005).

\bibitem[{Burrows} et~al.(1992)]{Burrows1992}
C.~{Burrows}, R.~{Burg}, and R.~{Giacconi}, \emph{\apj} \textbf{392}, 760
  (1992).

\end{thebibliography}

\IfFileExists{\jobname.bbl}{}
 {\typeout{}
  \typeout{******************************************}
  \typeout{** Please run "bibtex \jobname" to optain}
  \typeout{** the bibliography and then re-run LaTeX}
  \typeout{** twice to fix the references!}
  \typeout{******************************************}
  \typeout{}
 }

\end{document}